\documentstyle[11pt,newpasp,graphicx,twoside]{article}
\def\edcomment#1{\iffalse\marginpar{\raggedright\sl#1\/}\else\relax\fi}
\reversemarginpar

\begin{document}

\title{The Metallicity of Post-T Tauri Stars: A preliminary approach to the
understanding of the metal enrichment of stars harboring planets}
\author{Ramiro de la Reza, Licio da Silva and Roberta Almeida}
\affil{Observat\'orio Nacional, Rio de Janeiro - Brazil}
\author{Isa Oliveira}
\affil{Universidade Federal do Rio de Janeiro; Observat\'orio Nacional - Brazil}
\affil{isabel@pha.jhu.edu}
\author{Carlos Alberto O. Torres and Germano R. Quast}
\affil{Laboratorio Nacional de Astrof\'\i sica, Minas Gerais - Brazil}

\begin{abstract}

The metallicity of young low mass Post-T Tauri stars in coeval associations
is practically unknown. This work is the beginning of a systematic measurement 
of these metallicities based on high resolution spectra of low rotating members 
of these associations. Here, we present an application by examining the behavior 
of the Iron abundance with stellar mass and temperature of some members of an 
association 30 Myr old. This will test the possibility of explaining the high 
metallic content of stars with planets by means of injection of planetesimals 
during this early stage of evolution.
\end{abstract}

\section{Introduction}
A challenging and unexpected puzzle appeared with the discovery of exo-planets.
A large majority of stars with planets (SWP) present an important excess of
metallicity of the mean order of 0.25 dex (see review of Gonzales 2003).
At present time there is not a clear explanation for this. The problem can be
expressed this way: is metallicity the cause of planets, or are planets the cause
of metallicity? The investigations are essentially based on the two following
conceptions: {\it primordial explanation} - SWP are preferentially formed in
natal clouds already metal enriched; and {\it self-enrichment} produced by accretion
onto the star of different bodies deficient in H and He (e.g. asteroids, planets).
Due to the potential diversity of mechanisms involving the self-enrichment or pollution
processes, this explanation has received a large interest in the literature. The
primordial scenario appears today, only as an explanation when the self-enrichment
eventually fails. As we will see, the key of the self-enrichment process resides
in the behavior of the stellar convective layers. In fact, the small convective layer
of a hot and more massive F star will preserve the new incoming material and will 
therefore be more easily detected. This is not the case for a cooler and less massive
K star, in which the convective layer can be ten times larger. Other processes that 
can dilute the fresh material are mixing by rotation, mass loss, diffusion and 
gravitational settling; these are, however, less important than convection.

\section{Our approach}
The self-enrichment scenario has been invoked in the literature to occur
after the star has arrived on the main sequence (MS). At this stage of evolution the
star's convective layers have attained a stable minimum configuration. However, here
the accretion mechanism is likely to be inefficient, because few planetesimals are
expected to fall into the relatively old MS stars. One way to
solve this consists of invoking the infall of entire planets (this has been
mentioned several times in the literature with few details
except for the case of giant stars). This possibility seems, however, 
strange from a dynamical point of view. In fact, planets have  
stable orbits, even in the case of large eccentricities. Also, in this stage the 
stellar radius remains constant and the natural stellar mass loss, even modest will, 
on the contrary, provoke the planets to move away. 

Our approach is different. It consists of invoking a maximum fall of planetessimals
at the epoch when the latter were being formed and were probably accumulating to form 
terrestrial planets or the cores of giant planets. It is at that time that they were 
numerous and they  were able to bombard the star due to an effect produced by internal
migration of a young planet in the disk. This stage is supposed to be that of Post-T Tauri
stars with ages between 10 and 40 Myr. For this, we choose an association of $\sim$30 Myr
because it is at this age that stars with masses between 0.9 to $\sim$1.2 $M_\odot$,
which are the stars of interest here (see Figure 1), begin to establish their convective
layers (see for example D'Antona \& Mazzitelli, 1994). Our test will then consist  
of detecting the presence of the pollution signature in the form of an increase in 
metallicity for the hot and massive members of a coeval association. 

\subsection{The GAYA Association}
During our Southern Hemisphere survey SACY (Search for Associations Containing
Young stars) based on ROSAT X-ray sources, several associations were discovered
(Torres et al. 2003; Quast et al. 2003). Two of these are important in this context;
GAYA (Great Austral Young Association) with an age around 30 Myr and YSSA (Young
Scorpio-Sagitarius Association) with an age of $\sim$10 Myr. GAYA consists of two
structures: GAYA1 ($\sim$30 Myr) which contains the members of the known associations
Tucana (Zuckerman et al. 2001) and Horologium (Torres et al. 2000). GAYA 1 is
located at a mean distance of $\sim$45 pc. GAYA2, which appears to be somewhat younger 
($\sim$20 Myr), has a mean distance of $\sim$85 pc.

Here, we present the initial metallicity determinations of some low rotating members
of GAYA1 and GAYA2. Their projected rotation velocities vsin{\it i} are less than 40 km s$^{-1}$
corresponding to a mean equatorial velocity of $20\pm5$ km s$^{-1}$ (de la Reza \& Pinzon 
2004). Some very few members of YSSA are considered for comparison. To consider low rotation 
is important in order to avoid the blending of spectral lines, especially those of the 
scarce Fe II lines. All these measurements have been made using high resolution FEROS spectra 
having a large spectral coverage, obtained at the 1.52 m telescope at ESO.

\subsection{Methodology}
For each star, we have measured equivalent widths of nearly 50 Fe I lines and 11 lines of Fe II.
Calculations were performed by means of usual LTE procedures. A constant microturbulence
(see Table 1) has been maintained in all cases in order to disentangle the thermal effects.
In fact, temperature is a good indicator of surface convection in FGK stars
(Pinsonneault, DeFoy \& Coffee, 2001). Note that these spectral types are those of SWP.
The initial $T_{eff}$ values, before any iteration, have been obtained from observed $UBV(IR)_c$
photometry, using pre-MS models (Siess, Dufour \& Forestini, 2000) for the corresponding age. 
Some gravity values were obtained from Hipparcos parallaxes when available.

\begin{table}[h]
\caption{Main data of stars presented in Figure 1. Some stars are labeled GAYA because is not clear if they belong to GAYA1 or GAYA2}
\begin{tabular}{clclccccl}
\hline
\hline
N  & Name        & $T_{eff} $ & [Fe/H] & log $ g $ &  Li EW  & Micro & $ M/M_\odot $ & Assoc. \\
   &             &  (K)     &            &         &  (m\AA) &       &       &    \\
\hline
1  & HD47875     & 5833        & 0.28   & 4.3     & 210             & 2.0  & 1.2        & GAYA  \\
2  & HD987       & 5625        & 0.31   & 4.4     & 201             & 2.0  & 1.0        & GAYA1 \\
3  & T9217 0417  & 4675        & 0.085  & 4.2     & 330             & 2.0  & 0.9        & GAYA  \\
4  & HD81544     & 5437        & 0.05   & 4.4     & 340             & 2.0  & 1.1        & GAYA2 \\
5  & T8584 2682  & 5625        & 0.12   & 4.1     & 245             & 2.0  & 1.1        & GAYA2 \\
6  & HD26980     & 5773        & 0.13   & 4.2     & 183             & 2.0  & 1.2        & GAYA2 \\ 
7  & HD49855     & 5701        & 0.17   & 4.4     & 233             & 2.0  & 1.2        & GAYA2 \\
8  & HD222259    & 5714        & 0.14   & 4.4     & 231             & 2.0  & 0.9        & GAYA  \\
9  & HD55279     & 5373        & 0.019  & 4.4     & 275             & 2.0  & 1.0        & GAYA2 \\
10 & HD202917    & 5631        & 0.27   & 4.4     & 227             & 2.0  & 1.1        & GAYA1 \\
11 & T9243 1332  & 5833        & 0.15   & 4.3     & 220             & 2.0  & 1.1        & GAYA  \\
12 & HD274561    & 5250        & 0.15   & 4.0     & 270             & 2.0  & 0.9        & GAYA2 \\
13 & HD32372     & 5833        & 0.21   & 4.4     & 215             & 2.0  & 1.2        & GAYA2 \\
14 & T7044 0535  & 5185        & 0.01   & 3.9     & 300             & 2.0  & 1.1        & GAYA2 \\ 
15 & HD160682    & 5640        & 0.25   & 4.4     & 260             & 2.1  & 1.5        & YSSA  \\
16 & T7415 0284  & 5630        & 0.23   & 4.4     & 272             & 2.1  & 1.5        & YSSA  \\
\hline
\hline
\end{tabular}
\end{table}

\section{Results and Conclusions}
In Figure 1 and Table 1 we present the results of our measurements together with theoretical
enrichment models for MS stars by Pinsonneault et al. (2001). The three curves correspond to
accretion of 1, 3 and 10 Earth masses of Fe for an initial solar abundance. Similar curves for
initial values corresponding to [Fe/H] = +0.2 and [Fe/H] = $-0.2$ can be shifted up and down 
respectively. We notice a better adjustment if a solar initial value is chosen. This solar 
basis appears to be more appropriate to these young stars than the [Fe/H] = $-0.2$ value, 
which corresponds to the peak of the distribution of dwarfs in the solar neighborhood. Recently, 
Cody \& Sasselov (2004) studied the long term evolution of polluted stellar models without rotation.
The main effects are the expansion of the convective zone and downward shift of $T_{eff}$. In that
study the effects of surface pollution require significantly larger amounts of the injected material
than in the Pinsonneault et al. case. For instance, to replicate the 3 terrestrial masses of Fe in 
Figure 1, the calculations of Cody \& Sasselov require 40 Earth masses of Fe!  
In any case, these models were introduced for comparison to the metallicity of SWP with temperature
and no considerations were made on the eventual time variations of the pollution mechanisms.
In the case of Pinsonneault et al. they showed that accretion produces a much too high Fe abundance 
for stars hotter than 6000 K. The recent work of Cody \& Sasselov appears to have solved this point.
Also, following other considerations, Murray \& Chaboyer (2002) indicate that a smaller injection of
Fe of the order of 3 to 6 Earth masses can be compatible with SWP metallicities.

\begin{figure}[t]
\includegraphics[width=\textwidth]{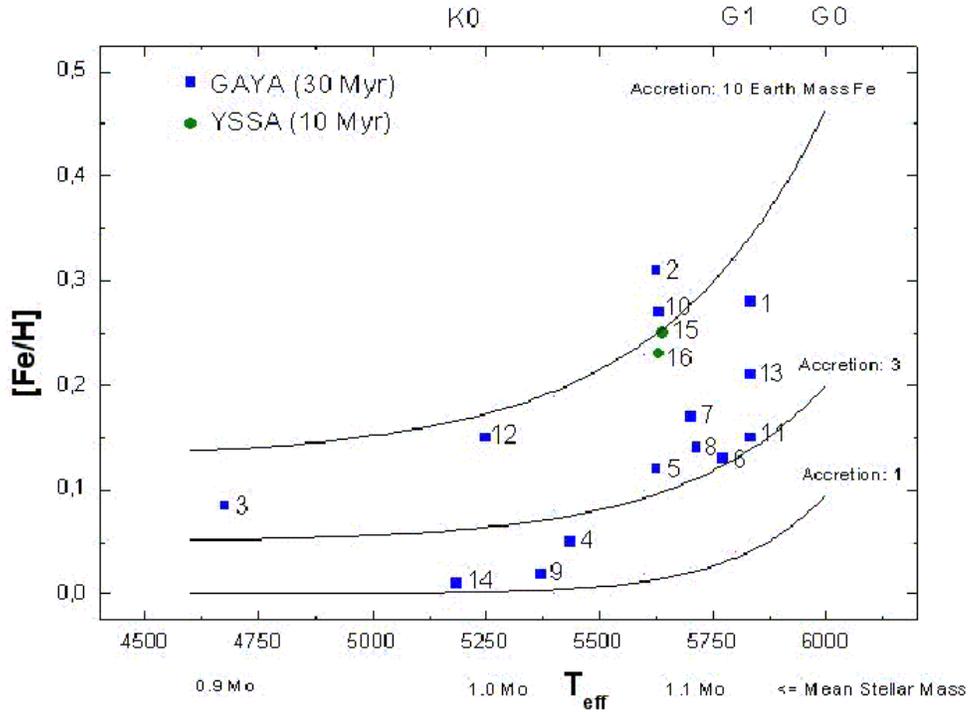}
\caption{Behavior of [Fe/H] abundances with stellar temperatures for some members of GAYA. Spectral
types and mean masses are shown. The label numbers correspond to those of Table 1. The curves are the MS
enrichment models of Pinsonneault et al. (2001) for three different Fe pollution accretions.}
\end{figure}

Even if this preliminary exploration seems to show a positive indication of the pollution process
at $\sim$30 Myr, we must confirm this trend by extending our measurements to more members of GAYA 
as well as field stars for comparison. Also, we note that a more detailed model must 
be constructed for the first 100 Myr of evolution to interpret our observations. Especially, this model 
must consider the dilution produced by rotation mixing. A first approach in this sense has been made by 
Pinsonneault et al. who obtained a dilution of 0.22 dex for a $\sim$6050 K star and 0.1 dex for a cooler star
at $\sim$5040 K produced by rotation with a mean velocity.

We will follow up on our measurements for other elements such as Li (see Table 1), Si and Ni, which are 
refractory elements as Fe, and CNO volatile elements. This will enable us to test the existence of ``hot''
(near the star) or cool accretion scenarios (Smith, Cunha \& Lazzaro, 2001).

\end{document}